\documentclass[conference]{IEEEtran}
\IEEEoverridecommandlockouts
\usepackage{cite}
\usepackage{amsmath,amssymb,amsfonts}
\usepackage{algorithmic}
\usepackage{graphicx}
\usepackage{textcomp}
\usepackage{xcolor}
\usepackage{multirow}

\usepackage{subfig}
\usepackage{caption}
\usepackage{booktabs}
\usepackage{algorithm}
\usepackage{algorithmic}
\usepackage{bm}
\usepackage{color}
\usepackage{transparent}
\usepackage{graphicx}
\usepackage{import}
\usepackage{amsmath}  
\def\BibTeX{{\rm B\kern-.05em{\sc i\kern-.025em b}\kern-.08em
    T\kern-.1667em\lower.7ex\hbox{E}\kern-.125emX}}
\begin{document}
	%
	\title{Underwater Differential Game: Finite-Time Target Hunting Task with Communication Delay}

	\author{\IEEEauthorblockN{Wei Wei\IEEEauthorrefmark{1}, JingJing Wang\IEEEauthorrefmark{2}, Jun Du\IEEEauthorrefmark{3}, Zhengru Fang\IEEEauthorrefmark{3}, Chunxiao Jiang\IEEEauthorrefmark{4}, and Yong Ren\IEEEauthorrefmark{3}}
		\IEEEauthorblockA{\IEEEauthorrefmark{1}Tsinghua Shenzhen International Graduate School, Tsinghua University, Shenzhen, 518055, China\\
			\IEEEauthorrefmark{2}School of Cyber Science and Technology, Beihang University, Beijing, 100191, China\\
			\IEEEauthorrefmark{3}Department of Electronic Engineering, Tsinghua University, Beijing, 100084, China\\
						\IEEEauthorrefmark{4}Tsinghua Space Center, Tsinghua University, Beijing, 100084, China\\
			Email: weiw20@mails.tsinghua.edu.cn, drwangjj@buaa.edu.cn, jundu@tsinghua.edu.cn, \\fangzr19@mails.tsinghua.edu.cn, jchx@tsinghua.edu.cn, reny@tsinghua.edu.cn}}
	
	%


	\maketitle
	
	\begin{abstract}
		This work considers designing an unmanned target hunting system for a swarm of unmanned underwater vehicles (UUVs) to hunt a target with high maneuverability. Differential game theory is used to analyze combat policies of UUVs and the target within finite time. The challenge lies in UUVs must conduct their control policies in consideration of
		not only the consistency of the hunting team but also escaping behaviors of the target. To obtain stable feedback control policies satisfying Nash equilibrium, we construct the Hamiltonian function with Leibniz’s formula. For further taken underwater disturbances and communication delay into consideration, modified deep reinforcement learning (DRL) is provided to investigate the underwater target hunting task in an unknown dynamic environment. Simulations show that underwater disturbances have
		a large impact on the system considering communication delay. Moreover, consistency tests show that UUVs  perform  better consistency with a relatively small range of disturbances.

	\end{abstract}
	

	%
	\IEEEpeerreviewmaketitle

	\section{Introduction}

	Recently, techniques on swarm intelligence focus on three main methods, model-based theory, Lyapunov analysis, and simulations. In comparison to model-based method, simulating approaches suffer from difficulties like convergence, accuracy, as well as complexity analyses. Moreover, Lyapunov analysis remains confined to boundary problems. 
	Model-based method, such as differential game theory, provides the proper framework to analyze conflicting interests of players involved in one or more swarm teams, and allows comprehensive theoretical analysis revealing structural properties.
	Therefore, target hunting tasks can be constructed with a differential game framework, where players are divided into opposite swarm teams: hunters and targets. Specifically, hunters perform tracking behaviors and finally encircle targets in their attacking scope, while targets prefer escaping from the searching area of hunters. 
	In \cite{8754740} and \cite{9122473}, authors studied differential game among multiple hunters and multiple targets,  and found the stable flying formation when reaching Nash equilibrium. 
	In \cite{7464902}, authors developed equilibrium open loop policies that discourage hunters from attacking, while encouraging retreating by solving the differential game of engagement.
	
	Despite the existing research in game-based target hunting area, few approaches have taken into consideration how dynamic environmental factors may affect the outperformance of differential game \cite{9451536}. On the one hand, the presence of sea currents or winds, respectively, may significantly affect the motion of a relatively small Unmanned Underwater Vehicle (UUV). As a result, during the underwater target hunting, optimal behaviors of UUVs, as  solutions to the differential game, may be greatly affected by the existence of external disturbances. Guidance laws in complex underwater environment should consider not only the maneuverability of players, but responses to disturbances \cite{fang2021}. 
	On the other hand, the difficulty encountered is the non-causality of control policies caused by communication delay. It shall be shown that the problem can be solved by introducing states that contain the past information and  capture future effects of control laws. 
	It just happens that along with batch learning, experience replay, and batch normalization, deep reinforcement learning (DRL) shows powerful capabilities to collect future effects and tackle complex tasks without much prior knowledge \cite{2019Human}. 
	Thus, in this paper, we construct a linear differential game model to analyze the underwater target hunting with a single target and multiple UUVs \cite{9422341}. Meanwhile, optimal feedback control policies can be obtained with the Hamiltonian function. 
	The challenge lies in UUVs must conduct their control policies in consideration of not only the consistency of swarm hunters, but also escaping behaviors of the target. In particular, UUVs must balance the competing objectives of keeping consistency, while avoiding collisions when catching the target. Further complications are caused by multi-stage hunting process, because UUVs must select appropriate control laws within each time slot to find feasible solutions of their respective. Moreover, modified DRL method is provided based on the differential game model to further investigate target hunting with communication delay and disturbances, while past information can be stored and future effects can be fed back to the current state.

	The remainder of this paper is outlined as follows. The system model and problem formulation are detailed to elaborate the  differential game with the underwater target hunting task in Section \uppercase\expandafter{\romannumeral2}. In Section \uppercase\expandafter{\romannumeral 3},
	the solution techniques are presented. In Section \uppercase\expandafter{\romannumeral 4},
	simulation results are provided for characterizing the proposed  differential game model with communication delay and disturbances, followed by conclusions in Section \uppercase\expandafter{\romannumeral 5}.
	
	%
	%
	
	
	%
	%
	
	
	%
	%
	%
	%
	%
	\section{System Model}
\begin{figure}[!t]
		
		\centering\includegraphics[width=8cm]{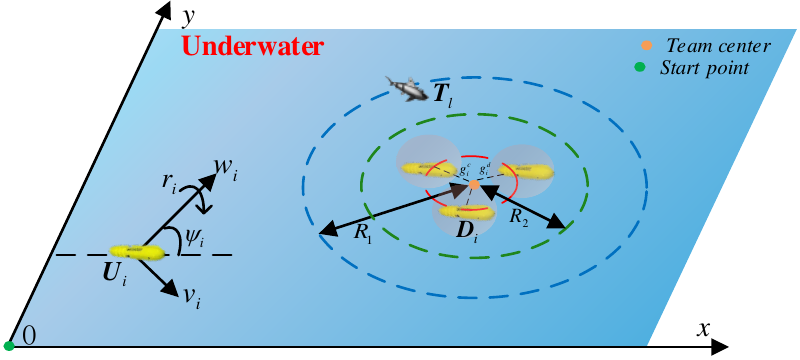}
		
		\centering\caption{Differential game between UUVs and the target.} 
		
		\label{fig:system}
		\vspace{-1.7em}
	\end{figure}

	In this section, we describe the system model, 
	assumptions, and some definitions used in this paper. Herein, an underwater multi-target hunting differential game is considered on a two-dimensional plane with the depth of $d$.
	The differential game is modeled to depict cooperative game among $M$ UUVs and non-cooperative game between UUVs and a single target.   
	As shown in Fig. \ref{fig:system}, the target $T_l$ is randomly distributed on a two-dimensional plane and UUVs' team $\bm{U}$ are dispersed around the start point $\bm O=(O_x,O_y,d)$ in the initial state. Their coordinates are defined as $\bm{T}_l=(t_{x_{l}}, t_{y_{l}}, d)$ and $\bm{U}=\cup _{i=1}^{M} \{\bm{U}_i=(u_{x_{i}}, u_{y_{i}}, d),\ i \in (0,M] \}$, respectively. Now, we have an assumption that UUVs and the target have perfect knowledge (speed, state, etc.) of each other, that is, the target hunting game is a perfect information differential game \cite{du2021oceanSHS}.
	\subsection{Dynamics of UUVs and the target}
	When a target is allocated, UUVs firstly tail after the target in the horizontal plane. Thus, we use a three-degrees-of-freedom underactuated UUV model with a body-fixed coordinate frame $\boldsymbol{v}_i=[ w_i,v_i,r_i ]^\text{T}$ and an earth-fixed reference frame $\bm\eta_i =[ u_{x_{i}}, u_{y_{i}},\psi_i ]^\text{T}$, where  $w_i$, $v_i$, and $r_i$ represent the surge, sway, and heave velocities \cite{8594674}. Besides, $\psi_i$ is the yaw angle. $\boldsymbol{v}_i$ is limited by the maximum speed $V_1$ satisfying $||\boldsymbol{v}_i|| \le V_1$. Then, the dynamics of the $i$-th UUV can be given by:
	\begin{equation}
		\dot{\bm{\eta}}_i=\bm J(\bm{\eta}_i) {\bm{v}_i},
		\label{dynamic1}
	\end{equation}
	\begin{equation}
		\bm{M}\dot{\bm{v}}_i+\bm{C}(\bm{v}_i) \bm{v}_i+\bm{B}(\bm{v}_i) \bm{v}_i+\bm{G}(\bm\eta_i)=\bm p_i+\bm\tau_d,
		\label{dynamic2}
	\end{equation}
	where $\bm{M}$ and $\bm{C}(\bm{v}_i)$ are the system inertia, including added mass, and the Coriolis-centripetal matrices, respectively. Moreover, $\bm{B}(\bm{v}_i)$ is the damping matrix and $\bm{G}(\bm{v}_i)$ is the resultant matrix of gravity and buoyancy. Herein, $\bm p_i$ is the control input, while $\bm\tau_d$ is the interference in the environment. 
	Besides, $\bm J(\bm{\eta}_i)$ is the transformation matrix which can be given by:
	\begin{equation}
		\boldsymbol{J}( \bm{\eta}_i) =\left[ \begin{matrix}
			\cos \psi_i&		-\sin \psi_i&		0\\
			\sin \psi_i&		\cos \psi_i&		0\\
			0&		0&		1\\
		\end{matrix} \right] .
	\end{equation}
	
	Similarly, we assume the velocity of the target $\boldsymbol{v}_T$ is limited by the maximum speed $V_2$, i.e. $||\boldsymbol{v}_T|| \le V_2$.
	Similarly,  the dynamics of the target is described with:
	\begin{equation}
		\dot{\bm{\eta}}_T=\bm J(\bm{\eta}_T) {\bm{v}_T},
		\label{dynamic3}
	\end{equation}
	\begin{equation}
		\bm{M}\dot{\bm{v}}_T+\bm{C}(\bm{v}_T) \bm{v}_T+\bm{B}(\bm{v}_T) \bm{v}_T+\bm{G}(\bm\eta_T)=\bm p_T+\bm\tau_d.
		\label{dynamic4}
	\end{equation}
	
	In the underwater target hunting game, UUVs show better chasing ability by fulfilling cooperation, thus we assume the acceleration of UUVs and the target satisfies $\|\dot {\boldsymbol{ v}}_i\|> \|\dot{\boldsymbol{ v}}_T\|$. As a single target performs more maneuverability than a team of UUVs when escaping, so we assume that the target has a wider range of movement, i.e. $\psi_i \in [ -\pi/2 ,\pi/2]$ and $\psi_T \in [ -\pi ,\pi]$.
	\subsection{Underwater Communication Delay}
	Generally, underwater communication delays are detrimental factors for UUVs and target to get the current knowledge of each other \cite{du2020learningVTM}.
	Meanwhile, the information exchange is based on the underwater acoustic transmission, and the speed of sounds in water can be calculated with the empirical formula:   
	\begin{equation}
		||\bm v_w||=1450+4.21T-0.037T^2+1.14(S-35)+0.175P,
	\end{equation}
	where $T$, $S$ and $P$ respectively represent the temperature, the salinity and the pressure \cite{5564139}. We assume  that acoustic waves travel in straight lines in underwater environments \cite{6464633}.
	
	Let $\bm{e}_i=\bm{T}_l-\bm{U}_i$ denote the position vector from the current position of the $i$-th UUV to the target's position. Thus, the communication delay from the $i$-th UUV to the target can be expressed as $\delta _{i\rightarrow T}=\boldsymbol{e}_i/(\boldsymbol{v}_T-\boldsymbol{v}_w)$, while the communication delay from the target to the $i$-th UUV can be expressed as $\delta _{T\rightarrow i}=\boldsymbol{e}_i/(\boldsymbol{v}_i-\boldsymbol{v}_w)$. The average communication delay $\delta$ can be expressed via the following formula:

	\begin{equation}
		\begin{aligned}
			\delta=\frac{1}{2M}\sum\nolimits_{i=1}^M\bigg({\frac{\boldsymbol{e}_i}{\boldsymbol{v}_T-\boldsymbol{v}_w}}+{\frac{\boldsymbol{e}_i}{\boldsymbol{v}_i-\boldsymbol{v}_w}}\bigg).
		\end{aligned}
	\end{equation}
	
	\subsection{Problem Formulation}
	We assume the searching range and the attacking range of each UUV are $R_1$ and $R_2$, respectively. The target can be detected by the $i$-th UUV when $\|\bm{e}_i\|<R_1$, and can be caught when $\|\bm{e}_i\|<R_2$. 
	The multi-UUV cooperative target hunting problem consists in determining feedback control strategies $\bm{\eta}_i$ and $\bm{v}_i$, that steer each UUV from its initial position to the target, while avoiding collisions and being too far away from other UUVs, and thus maintaining consistency of the team.
	
	Consider a dynamic game model for the time evolution as described by the ordinary differential equation \cite{6922516}, we define the state function between UUVs and a single target as:
	\begin{equation}
		\begin{aligned}
			\boldsymbol{\dot{s}}(t)&=\bm F_s\boldsymbol{s}(t-\delta)+\bm G_{12}\boldsymbol{p}(t)+\bm G_{21}\boldsymbol{q}(t) , \ t\in[0,\ T_h],\\
			\boldsymbol{s}\left( 0 \right) &=\boldsymbol{s}_0,\ t\in[-\delta,\ 0],
		\end{aligned}
		\label{systemm}
	\end{equation}
	where $\bm s(t)=[\bm{U}_{1}^{\text{T}}(t),\bm{U}_{2}^{\text{T}}(t),...,\bm{U}_{M}^{\text{T}}(t),\bm{T}_l^{\text{T}}(t)]^{\text{T}} \in R^{n\times 1}$  ($n=M+1$) represents the joint position configuration of $M$ UUVs and the target  at time $t$. Moreover, $\bm p_i(t)=[\dot w_i(t),\dot v_i(t),\dot r_i(t),\psi_i(t)]^{\text{T}}$ stands for the control input of the $i$-th UUV. 
	Then, control inputs of UUVs with choices of speeds and headings can be further expressed as $\bm p(t)=[\bm{p}_{1}^{\text{T}}(t),\bm{p}_{2}^{\text{T}}(t),...,\bm{p}_{M}^{\text{T}}(t)]^{\text{T}} \in R^{m}$, while $\bm q(t)=[\dot w_T(t),\dot v_T(t),\dot r_T(t),\psi_T(t)]^\text{T}$ is the control input of the target. Besides, $T_h$ is the maximum hunting time. Moreover, $\bm{s}_0$ is the initial condition, while $\bm{s}_f$ is the final condition when $\|\bm{e}_i(t)\|>R_1$ or $\|\bm{e}_i(t)\|<R_2$.
	Furthermore, $\bm F_s \in R^{n\times n}$, $\bm G_{12} \in R^{n\times m}$, and $\bm G_{21} \in R^{n\times 4}$ are coefficient matrices \cite{6489503}.

	We consider a multi-UUV system consisting of $M$ UUVs with dynamics (1) and (2), for $i \in (0,M]$, and let $\boldsymbol{e}(t)=\left[ \boldsymbol{e}_{1}^{\text{T}}(t),\boldsymbol{e}_{2}^{\text{T}}(t),...,\boldsymbol{e}_{M}^{\text{T}}(t) \right] ^{\text{T}}$. Thus, the pay-off function of the $i$-th UUV  with constants $\alpha _{i}^{d}>0$, $\beta _{i}^{c}>0$ can be expressed by:
	\vspace{-1em}
	\begin{equation}
		\begin{aligned}
			P_i(\boldsymbol{p}_i,\boldsymbol{q},\boldsymbol{s}_0)=&\frac{1}{2}\int_0^{T_h}{\bm{p}_{i}^{\text{T}}\left( \alpha _{i}^{d}g_{i}^{d}+\beta _{i}^{c}g_{i}^{c} \right) \bm{p}_i }dt
			-\bm{s}_f^\text{T}\phi_i(\bm{s}_f)\bm{s}_f.
		\end{aligned}
		\label{pay-off-UUV}
	\end{equation}

	We define that each UUV has a safety radius $r$ to avoid collisions with other UUVs.
	The $i$-th UUV is said to collide with the $j$-th UUV if there exists a time instant $t$ such that $||\bm{U}_i(t)-\bm{U}_j(t)||^2 \le r$, for $i \in (0,M]$ and $j \in (0,M]$. 
	Then, we define the UUV avoidance region of the $i$-th UUV at $t$ as $D_i=\cup _{j=1,j\ne i}^{M}D_{ij}$, where $D_{ij}=\{||\bm{U}_i(t)-\bm{U}_j(t)||^2 \le r,j \in (0,M], j\ne i\}$.
	Thus, $g_{i}^{d}(t)$ can be defined as the function penalizing the $i$-th UUV from approaching other UUVs, hence can be considered as collision avoidance function:
	\begin{equation}
		g_{i}^{d}(t)=\sum\nolimits_{j=1,j\ne i}^N{\left( \lVert \boldsymbol{U}_i(t)-\boldsymbol{U}_j(t) \rVert ^2-r^2 \right) ^{-c}},
	\end{equation}
	where $c>0$ and ${\lim}_{U_i\rightarrow \partial D_i}g_{i}^{d}=+\infty$.
	
	Consistency is a key technology for multiple UUVs to coordinate and cooperate with each other to complete complex hunting tasks. UUVs are considered to realize the consistency such that
	${\lim}_{t\rightarrow \infty}\ \lVert U_i\left( t \right) -U_j\left( t \right) \rVert =0\ (\forall i,j=1,2,...,M)$
	holds under any initial conditions \cite{8506620}. Thus, $g_{i}^{c}(t)$ can be defined as the function penalizing the $i$-th UUV away from other UUVs, hence can be considered as consistency functions:
	\begin{equation}
		g_{i}^{c}(t)=\sum\nolimits_{j=1,j\ne i}^M{ \lVert \boldsymbol{U}_i(t)-\boldsymbol{U}_j(t) \rVert ^2 }.
	\end{equation}
	
	When the target enters UUVs' attacking range with radius $R_2$  or escapes from UUVs' sensing range with radius $R_1$, the differential hunting game ends. Thus, the terminal value function with constants $a$ and $b$ at  $\bm s_f$ is defined as:
	\begin{equation}
		\phi_i(\bm s_f)=\left\{ \begin{array}{l}
			1/a,\ \|\bm{e}_i\|>R_1,\\
			1/b,\ \|\bm{e}_i\|<R_2.\\
		\end{array} \right. 
	\end{equation}
	
	To avoid being captured, the selfish target has three goals: (1) maximizing the distance to UUVs to avoid being chased; (2) minimizing its own control effort; and (3) maximizing the control effort $\lVert \boldsymbol{q}(t) \rVert$ of UUVs, such that UUVs would take more efforts to hunt the target and the target would have more chances to flee away.
	Thus, the pay-off function of the target related to the $i$-th UUV can be expressed as:
	\begin{equation}
		\begin{split}
			P_T^i(\boldsymbol{p}_i,\boldsymbol{q},\boldsymbol{s}_0)=\frac{1}{2}\int_0^{T_h}{\bm{q}^{\text{T}}\bigg(\frac{1}{ \lVert \boldsymbol{U}_i(t)-\boldsymbol{T}_l (t) \rVert ^2 }  \bigg)\bm{q}}dt.
			\label{pay-off-target}
		\end{split}
	\end{equation}
	
	In general, the pay-off function of the underwater target hunting system can be designed according to (\ref{pay-off-UUV}) and (\ref{pay-off-target}):
	\begin{equation}
		\begin{aligned}
			&P_E(\boldsymbol{p},\boldsymbol{q},\boldsymbol{s}_0)
			=\sum\nolimits_{i=1}^M{\bigg\{P_i\left( \boldsymbol{p}_i,\boldsymbol{q,s}_0 \right)}-P_T^i(\boldsymbol{p}_i,\boldsymbol{q},\boldsymbol{s}_0)\bigg\}\\
			&=\frac{1}{2}\sum\nolimits_{i=1}^M\bigg\{\int_0^{T_h}{\bm{p}_{i}^{\text{T}}\left( \alpha _{i}^{d}g_{i}^{d}(t)+\beta _{i}^{c}g_{i}^{c}(t) \right) \bm{p}_i  }dt\\
			&-\int_0^{T_h}{\bm{q}^{\text{T}}\big[ \lVert \boldsymbol{U}_i(t)-\boldsymbol{T}_l (t) \rVert ^{-2}   \big]}\bm{q}dt-\bm{s}_f^\text{T}\phi_i(\bm{s}_f)\bm{s}_f\bigg\}.
		\end{aligned}
		\label{pay-off-system}
	\end{equation}
	
	Using the pay-off function (\ref{pay-off-system}), we define the differential game where each participant attempts to minimize their respective pay-off functions for a given initial state \cite{7464902}, which can be expressed as:
	\begin{equation}
		P_{E}^{*}\left(\bm q,\bm s_0 \right) =\underset{\boldsymbol{p}}{\min}\ P_E(\boldsymbol{p},\boldsymbol{q},\boldsymbol{s}_0).
	\end{equation}
	Moreover, the equilibrium value $P_{E}^{*}\left( \boldsymbol{s}_0 \right)$ satisfies the following Nash equilibrium condition:
	\begin{equation}
		P_{E}\left( \bm p^*,\bm q,\bm s_0 \right) \le P_{E}\left( \bm p^*,\bm q^*,\bm s_0 \right) =P_{E}^*(\bm s_0) \le P_{E}\left( \bm p,\bm q^*,\bm s_0 \right).
	\end{equation}
	
	\emph{Problem definition}: Taken the system function (\ref{systemm}) and pay-off function (\ref{pay-off-system}) into consideration, solving the UUV-target, non-cooperative underwater target hunting differential game consists in determining an admissible pair of feedback strategies ($\bm p^*$,$\bm q^*$) such that $	P_{E}\left( \bm p^*,\bm q,\bm s_0 \right) \le P_{E}\left( \bm p^*,\bm q^*,\bm s_0 \right) =P_{E}^*(\bm s_0) \le P_{E}\left( \bm p,\bm q^*,\bm s_0 \right)$. Moreover, UUVs always terminate the game in catching the target, while minimizing their respective pay-off functions. Simultaneously, the target attempts to maximize UUV’s pay-off function throughout the course of differential game. Using these goals along with dynamics of UUVs and the target, the differential game is defined as:
	\begin{equation}
		V_E^*(\bm s_0):=\underset{\boldsymbol{p}}{\min}\ \underset{\boldsymbol{q}}{\max}\left\{ P_E (\boldsymbol{p},\boldsymbol{q},\boldsymbol{s}_0)\right\} 
		\label{17},
	\end{equation}
	subject to (\ref{dynamic1}), (\ref{dynamic2}), (\ref{dynamic3}), (\ref{dynamic4}), with the final condition $\bm{s}_f$  if there exists an UUV $i$ satisfying $\|\bm{e}_i\|>R_1$ or $\|\bm{e}_i\|<R_2$.

	\section{Solution Technique}
	In this section,   the Hamiltonian function is used to gain feedback control policies. Meanwhile, the modified DQN method is further proposed to study the influence of delay and disturbances on target hunting differential game.

	\subsection{Optimal Control Policies for Underwater Target Hunting with $\delta=0$}
	The function $V_E(\bm s_0)$ represents the equilibrium value of the
	game starting at $\boldsymbol{s}_0$ when UUVs and the target implement their respective equilibrium control strategies $\bm p^*$ and $\bm q^*$, which can be solved by:
	\begin{equation}
		\bm p^*,\bm q^*=\text{arg}\ \underset{\boldsymbol{p}}{\min}\ \underset{\boldsymbol{q}}{\max}\left\{ P_E (\boldsymbol{p},\boldsymbol{q},\boldsymbol{s}_0)\right\} .
	\end{equation}
	Given the feedback policy pair $(\bm p, \bm q)$, the cost of policy pair at time $t$ such as $V_E(\bm s(t))$ can be defined as \cite{7332778}:
	\begin{equation}
		\begin{aligned}
			&V_E(\bm s(t))=\frac{1}{2}\sum\nolimits_{i=1}^M\bigg\{\int_t^{T_h}{\bm{p}_{i}^{\text{T}}\left( \alpha _{i}^{d}g_{i}^{d}(t)+\beta _{i}^{c}g_{i}^{c}(t) \right) \bm{p}_i  }dt\\
			&-\int_t^{T_h}{\bm{q}_{i}^{\text{T}}\big[\lVert \boldsymbol{U}_i(t)-\boldsymbol{T}_l (t) \rVert ^{-2}  \big]}\bm{q}_{i}dt-\bm{s}_f^\text{T}\phi_i(\bm{s}_f)\bm{s}_f\bigg\}.
			\label{V_E}
		\end{aligned}
	\end{equation}
	
	Since the value of (\ref{V_E}) is finite, a differential equivalent can be found by using Leibniz’s formula and differentiating. Thus, the Hamiltonian function can be constructed as \cite{7464902}:
	\begin{equation}
		\begin{aligned}
			0=&\nabla{V}_E^{\text{T}}\cdot\boldsymbol{\dot{s}}(t) + \frac{1}{2}\sum\nolimits_{i=1}^M\bigg\{\bm{p}_{i}^{\text{T}}\left( \alpha _{i}^{d}g_{i}^{d}(t)+\beta _{i}^{c}g_{i}^{c}(t) \right) \bm{p}_i \\
			&-\bm{q}^{\text{T}}\frac{1}{ \lVert \boldsymbol{U}_i(t)-\boldsymbol{T}_l (t) \rVert ^2 }\bm{q}  \bigg\}
			:=H_E(\boldsymbol{s},\boldsymbol{p},\boldsymbol{q},\triangledown{V}_E),
		\end{aligned}
		\label{HE}
	\end{equation}
	where $\nabla{V}_E=\left( \partial V_E/\partial_{\bm{U}_1},\cdots, \partial V_E/\partial_{\bm{U}_M},\partial V_E/\partial_{\bm{T}_l}\right)^{\text{T}}$. 
	
	%
	%
	%
	%
	
	Thus, the optimal control strategies for UUVs and the target are found by maximizing or minimizing the Hamiltonian function appropriately. Furthermore, the necessary condition for the Nash condition in (\ref{17}) can be reformulated as:
	\begin{equation}
		\bm p^*,\bm q^*=\text{arg\ }\underset{\boldsymbol{p}}{\min}\ \underset{\boldsymbol{q}}{\max}\left\{ H_E(\boldsymbol{s},\boldsymbol{p},\boldsymbol{q},\nabla{V}_E)\right\},
	\end{equation}
	for each feedback policy pair $(\bm p, \bm q)$. When reaching Nash equilibrium, there exist stationary conditions:
	\begin{subequations}
		\begin{align}
			&\partial H_E/\partial \bm s= \bm F_s^{\text T}\nabla{V}_E=-\nabla^2{V}_E, \label{Za}\\
			&\partial H_E/\partial  \nabla{V}_E=\bm F_s\boldsymbol{s}+\bm G_{12}\boldsymbol{p}+\bm G_{21}\boldsymbol{q}=\bm{\dot{s}},\label{Zb}\\
			&\frac{\partial H_E}{\partial \bm p_i}=\left( \alpha _{i}^{d}g_{i}^{d}(t)+\beta _{i}^{c}g_{i}^{c}(t) \right)\bm{p}_i  +\bm G_{12}^{\text{T}} \nabla V_E^*=0,\label{Zc}\\
			&\frac{\partial H_E}{\partial \bm q}=-\sum_{i=1}^M{ \lVert \boldsymbol{U}_i(t)-\boldsymbol{T}_l (t) \rVert ^{-2} }\bm{q}+\bm G_{21}^{\text{T}} \nabla V_E^*=0\label{Zd}.
		\end{align}
	\end{subequations}
	
	\begin{figure}[!t]
		\centering
		\includegraphics[width=0.7\linewidth]{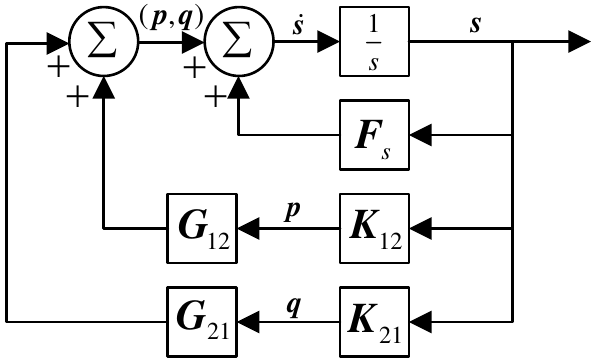}
		
		\centering\caption{The feedback control diagram of underwater target hunting game.} 
		\label{fig:controlfeedback}
	\end{figure}
	
	Furthermore, optimal control policies $\bm p_i^*\left( \bm s \right)$ and $\bm q^*\left( \bm s \right)$, which set up for all UUVs and the target, can be obtained by jointly applying (\ref{Zc}) and (\ref{Zd}) to (\ref{HE}):
	\begin{equation}
		\begin{aligned}
			\bm p_i^*\left( \bm s \right) &=-\big[\alpha _{i}^{d}g_{i}^{d}(t)+\beta _{i}^{c}g_{i}^{c}(t)\big]^{-1} \bm G_{12}^{\text{T}}\left( s \right) \nabla V_E^*\left( s \right) ,\\
			\bm q^*\left( \bm s \right) &=\frac{ \bm G_{21}^{\text{T}}\left( s \right) \nabla V_E^*\left( s \right)}{\sum_{i=1}^M{\lVert \boldsymbol{U}_i(t)-\boldsymbol{T}_l (t) \rVert ^{-2} }  }.
			\label{optimal-control}
		\end{aligned}
	\end{equation}

	Accordance with the  assumption $\nabla{V}_E=\bm P \cdot \bm s\ (\bm P \in R^{n\times n})$,
	the equation in (\ref{optimal-control}) can be reformulated as:
	\begin{equation}
		\begin{aligned}
			\bm p_i^*\left( \bm s \right) &=-\big[\alpha _{i}^{d}g_{i}^{d}(t)+\beta _{i}^{c}g_{i}^{c}(t)\big]^{-1} \bm G_{12}^{\text{T}}\left( s \right) \bm P  \bm s ,\\
			\bm q^*\left( \bm s \right) &=\frac{ \bm G_{21}^{\text{T}}\left( s \right) \bm P  \bm s}{\sum_{i=1}^M{\lVert \boldsymbol{U}_i(t)-\boldsymbol{T}_l (t) \rVert ^{-2} }  }.
		\end{aligned}
		\label{control-new}
	\end{equation}

	Applying $\nabla{V}_E=\bm P \cdot \bm s$ to the stationary condition (\ref{Za}), we gain the equation $\bm F_s^{\text T}  \bm P  \bm s+\bm {\dot{P}}  \bm s+\bm P  \bm {\dot{s}}=0$. Furthermore, Riccati equation (\ref{MRDE}) can be obtained by applying (\ref{control-new}) to (\ref{systemm}).
	Specifically, for $\forall \bm s$, there exists a symmetric matrix $\bm P$ with the terminal state function $\bm P_f=\bm s_f$ satisfying:
	\begin{equation}
		\begin{aligned}
			&\bm F_s^{\text T}  \bm P  +\bm {\dot{P}}  +\bm P  \bm F_s
			-\bm P\bm G_{12}\sum_{i=1}^M\big[\alpha _{i}^{d}g_{i}^{d}(t)+\beta _{i}^{c}g_{i}^{c}(t)\big]^{-1} \bm G_{12}^{\text{T}} \bm P \\
			&	+\bm P\bm G_{21}\frac{ \bm G_{21}^{\text{T}}\left( s \right) \bm P  }{\sum_{i=1}^M{\lVert \boldsymbol{U}_i(t)-\boldsymbol{T}_l (t) \rVert ^{-2} }  }  =0.
			\label{MRDE}
		\end{aligned}
	\end{equation}
	
	
	The feedback control laws can be expressed as $\bm p=\bm K_{12}\bm s$ and $\bm q=\bm K_{21}\bm s$, while feedback constraints $\bm K_{12}=[{\bm{K}_{12}^1}^{\text{T}},{\bm{K}_{12}^2}^{\text{T}},...,{\bm{K}_{12}^M}^{\text{T}}]^{\text{T}}$ and $\bm K_{21}$ shown in Fig. \ref{fig:controlfeedback} respectively satisfying:
	\begin{equation}
		\begin{aligned}
			\bm K_{12}^i&=-\big[\alpha _{i}^{d}g_{i}^{d}(t)+\beta _{i}^{c}g_{i}^{c}(t)\big]^{-1} \bm G_{12}^{\text{T}}\left( s \right) \bm P,\\
			\bm K_{21}&=\frac{ \bm G_{21}^{\text{T}}\left( s \right) \bm P }{\sum_{i=1}^M{\lVert \boldsymbol{U}_i(t)-\boldsymbol{T}_l (t) \rVert ^{-2} }  }.  
		\end{aligned}
	\end{equation}

	\subsection{DQN-Based Algorithm for Underwater Target Hunting with $\delta\in [0,T_h-t_0]$}
	Deep Q-learning (DQN), as an important component of DRL, can store past information and pass back future effects through target network \cite{8957702}. By using DQN, we can jointly simulate the communication delay and underwater distributions \cite{2019Human}. Along with real-time state $\bm s(t)$ and control policies [$\bm p(t)$, $\bm q(t)$], the reward function that motivates UUVs to complete the target hunting task is negative to pay-off function (\ref{pay-off-system}) in section \uppercase\expandafter{\romannumeral 2}, which can be further expressed as:
	\begin{equation}
		\begin{aligned}
			R_E =\left\{ \begin{array}{l}
				R_E+1/{P_E\left(\bm p, \bm q, \bm s(t-\delta) \right)}, \forall \ \|\bm{e}_i\|\in [R_2,R_1],\\ 
				a,\ \exists \ \|\bm{e}_i\|>R_1,\\
				b,\ \exists \ \|\bm{e}_i\|<R_2.\\
			\end{array} \right. 
		\end{aligned}
		\label{reward}
	\end{equation} 
	
	Herein, Q-value ($P_E^*$) will be iteratively updated when UUVs conduct control policies, which can be provided by: 
	\begin{equation}
		\begin{aligned}
			P_E^*={\text E}_{\bm s^\prime \sim \bm s}[ r+\chi \underset{(\bm p^\prime, \bm q^\prime)}{\min}P_E\left(\bm s^\prime,\bm p^\prime, \bm q^\prime \right) |\bm s(t-\delta),\bm p, \bm q ],
		\end{aligned}
		\label{Qvalue}
	\end{equation}
	where $0 \leq \chi  \leq 1$ is a discounting factor to decrease the weight of future rewards, while $\bm s^\prime$, $\bm p^\prime$, and $\bm q^\prime$ are the state and policies in the next time slot. The process to accomplish the underwater target hunting task with communication delay and disturbance is illustrated in Algorithm 1.

	\begin{algorithm}[!t]
		\renewcommand{\algorithmicrequire}{\textbf{Input:}}
		\renewcommand{\algorithmicensure}{\textbf{Output:}}
		\caption{DQN-Based Algorithm for Underwater Target Hunting Task with communication delay $\delta\in [0,T_h-t_0)$.}
		\label{alg1}
		\begin{algorithmic}[1]
			\REQUIRE 
			Admissible control policy ($\bm p_0$, $\bm q_0$), initial state $\bm s_0$, disturbances $\bm\tau_d$, $V_E=0$, time slot $k=0$.	
			\REPEAT
			\STATE 
			Calculate the underwater communication delay:
			\begin{equation}
				\delta=\bigg\lfloor \frac{1}{2M}\sum_{i=1}^M\bigg({\frac{\boldsymbol{e}_i}{\boldsymbol{v}_T-\boldsymbol{v}_w}}+{\frac{\boldsymbol{e}_i}{\boldsymbol{v}_i-\boldsymbol{v}_w}}\bigg) \bigg\rfloor.
			\end{equation}	
			
			\IF {$k-\delta \le t_0$}
			\STATE Randomly waking with initial settings.	
			\ELSE 
			\STATE Given state $\boldsymbol{s}(k-\delta)$, update control policies $\bm p(k)$, $\bm q(k)$ based on $P_E^*$.
			\STATE Calculate reward $R_E$ based on (\ref{reward}).
			\STATE Calculate Q-value $P_E^*$ based on (\ref{Qvalue}).
			
			\ENDIF 
			\STATE $k \leftarrow k + 1$.
			\UNTIL $\|\bm{e}_i(k)\|>R_1$ or $\|\bm{e}_i(k)\|<R_2$.  
			\ENSURE  $\boldsymbol{\dot{s}}(t)$. 
			\label{Algorithm1}
		\end{algorithmic}  
	\end{algorithm} 

	\section{Simulation Results} 
	\addtolength{\topmargin}{0.01in}
	In our simulations,  UUVs' hunting center $\bm O$ is located in $(400, 400, -200)$ initially with $M=3$ UUVs. The target is randomly distributed 40 m away from UUVs, i.e. $\|\bm{T}_l-\bm O\|$= 40 m.
	As for the speed of players, UUVs and target both begin with 1 knot\footnote{1 knot=1.852 km/h}, while the maximum speed of UUVs limited by $V_1=5$ knot and the maximum speed of target limited by $V_2=1$ knot. Herein, we assume UUVs and the target have a stable acceleration at each time slot with $\psi_i \in [ -\pi/2 ,\pi/2]$ and $\psi_T \in [ -\pi ,\pi]$.
	Moreover, the safe radius, the sensing radius of UUV, the attacking radius of UUV are set to 5 m, 80 m and 15 m, respectively. Since the target hunting task is finite-time, maximum time slots during one episode are set to 1000. Besides, constraints a, b, c are used with -1, 10 and 0.5, respectively. 
	We implement the modified DQN with Pytorch and conduct 5000 episodes experiments to verify the performance, where the structure of DQN is established with a fully connected
	neural network including two hidden layers. 
    Table \ref{22} shows parameters of the system and  Algorithm 1.
	\begin{table}[t!]
		\renewcommand\arraystretch{0.8}
		\newcommand{\tabincell}[2]{\begin{tabular}{@{}#1@{}}#2\end{tabular}}%
		\centering
		\caption{Parameters of System and Algorithm}
		\begin{tabular}{c|l|r}
			\toprule
			\toprule
			& \textbf{Parameters} & \textbf{Values} \\
			\midrule
			\multirow{9}[2]{*}{\tabincell{c}{\textbf{DQN}\\\textbf{parameters}}} & Learning rate ($\xi$)      &  0.0002\\
			& Training episodes ($ \ell$)  & $5000$ \\
			& Discounting factor ($\chi$)     & 0.9 \\
			& Batch size      & 128 \\
			& Memory capacity      & 10000 \\
			& $\epsilon$-greedy ($\epsilon$)     & 0.9 \\
			\midrule
			\multirow{13}[2]{*}{\tabincell{c}{\textbf{System}\\\textbf{parameters}}}
			& Start point of UUVs ($\bm O)$& (400, 400, -200) m\\
			& Number of UUVs ($M$) & 3 \\
			& Initial distance $\|\bm{T}_l-\bm O\|$ & 40 m \\
			& Maximum speed  of UUV ($V_1$)& 5 knot \\
			& Maximum speed  of target ($V_2$)& 2 knot \\
			& Acceleration of UUV ($\|\dot{\bm{ v}_i}\|$)& 0.008 knot/s \\
			& Acceleration of target ($\|\dot{\bm{ v}_T}\|$)& 0.0016 knot/s \\
			& movement range of UUV ($\psi_i$)& $[ -\pi/2 ,\pi/2]$ \\
			& Movement range of target ($\psi_T$)& $[ -\pi ,\pi]$ \\
			& Safe radius of UUV ($r$) & 5 m \\
			& Sensing  radius of UUV ($R_1$) & 80 m \\
			& Atattcking radius of UUV ($R_2$) & 15 m \\
			& Initial speed of UUVs ($V_G$) & 1 knot \\
			& Initial speed of the target ($V_t$)& 1 knot \\
			& Maximum number of time slots & 1000 \\
			& Constraint ($a$) & -1 \\
			& Constraint ($b$) & 10 \\
			& Constraint ($c$) & 0.5 \\
			\bottomrule
		\end{tabular}%
		\label{22}%
	\end{table}%

	\begin{figure*}[t!]
		\centering\subfloat[Underwater paths when $\delta=0$.] 
		{
			\begin{minipage}[t]{6cm}
				\centering
				\includegraphics[height=4.5cm]{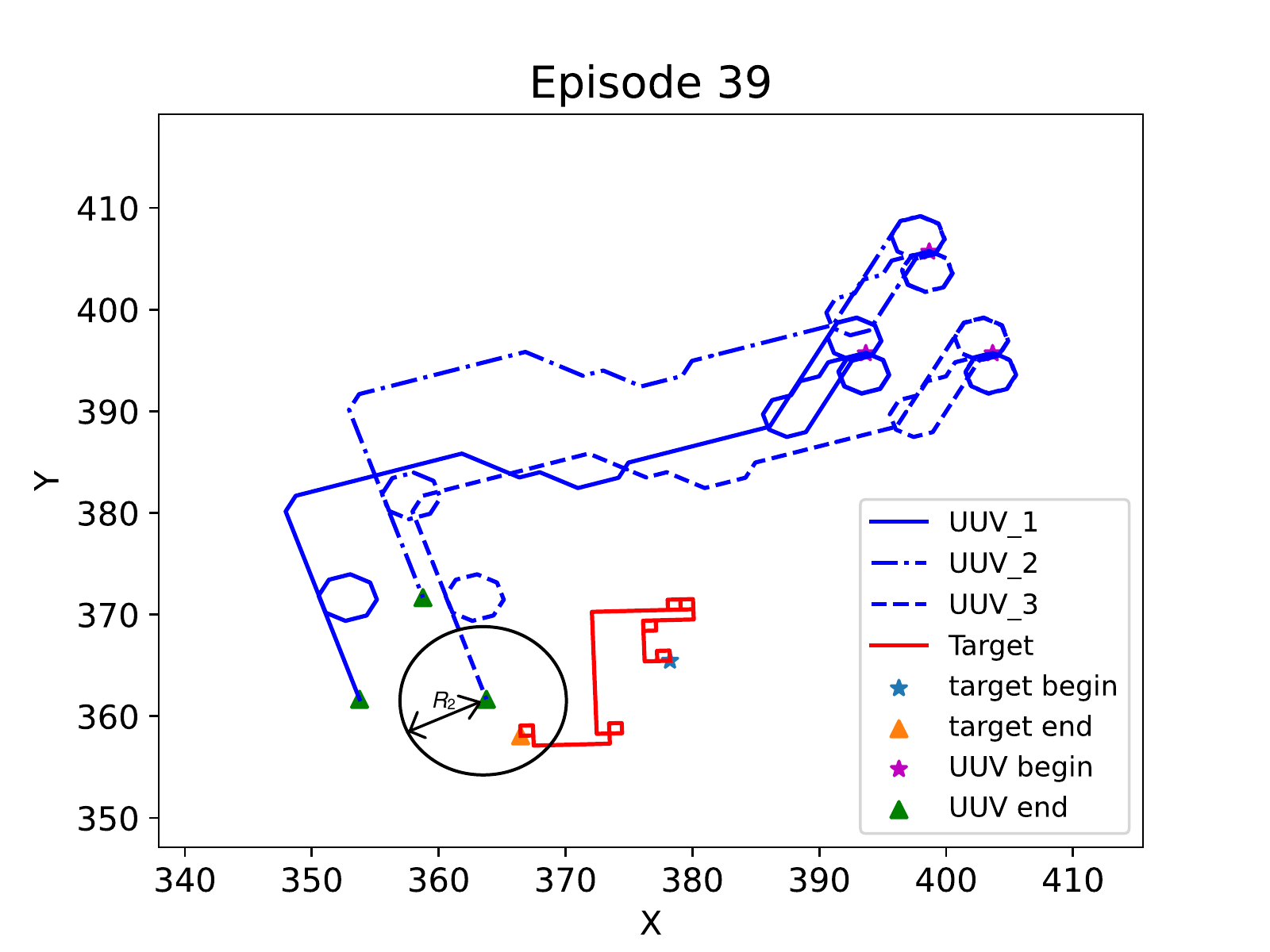}
			\end{minipage}%
		}
		\centering\subfloat[Rewards with various $\bm \tau_d$ when $\delta=0$.] 
		{
			\begin{minipage}[t]{6cm}
				\centering
				\includegraphics[height=4.5cm]{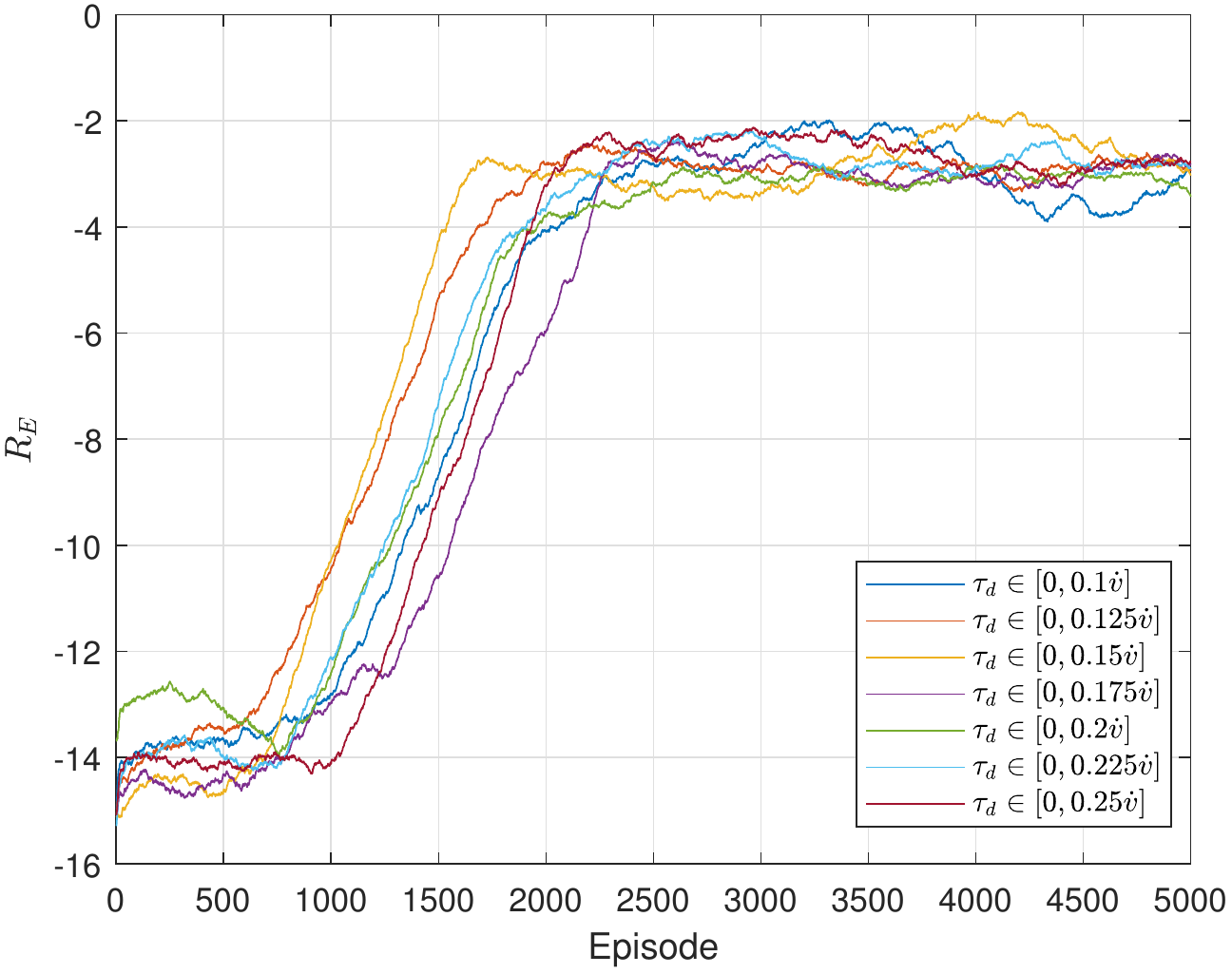}
			\end{minipage}%
		}
		\centering\subfloat[Consistency test with various $\bm \tau_d$ when $\delta=0$.] 
		{
			\begin{minipage}[t]{6cm}
				\centering
				\includegraphics[height=4.5cm]{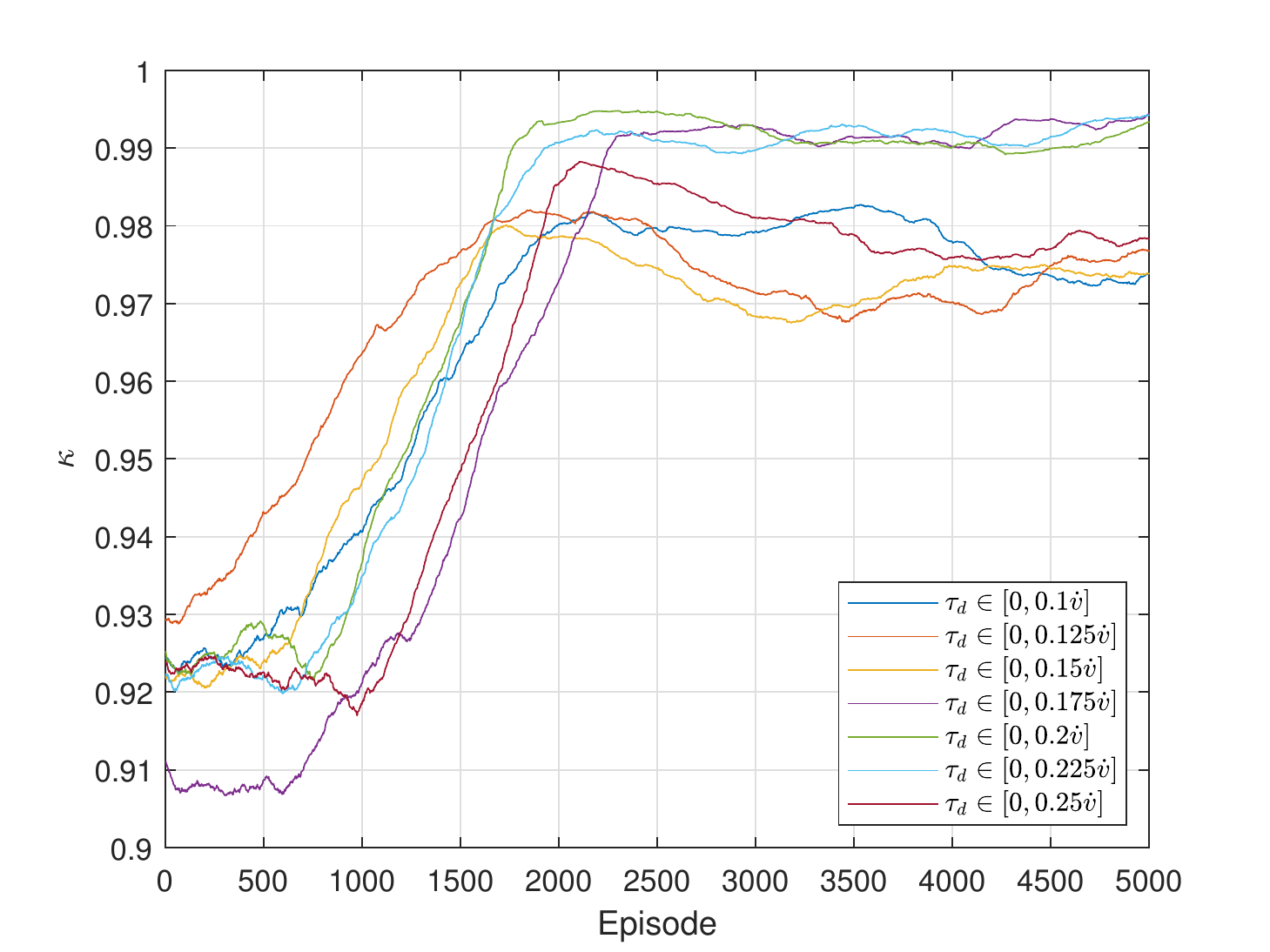}
			\end{minipage}%
		}
		
		\centering\subfloat[Underwater paths when $\delta \in (0,T_h-t_0)$.] 
		{
			\begin{minipage}[t]{6cm}
				\centering
				\includegraphics[height=4.5cm]{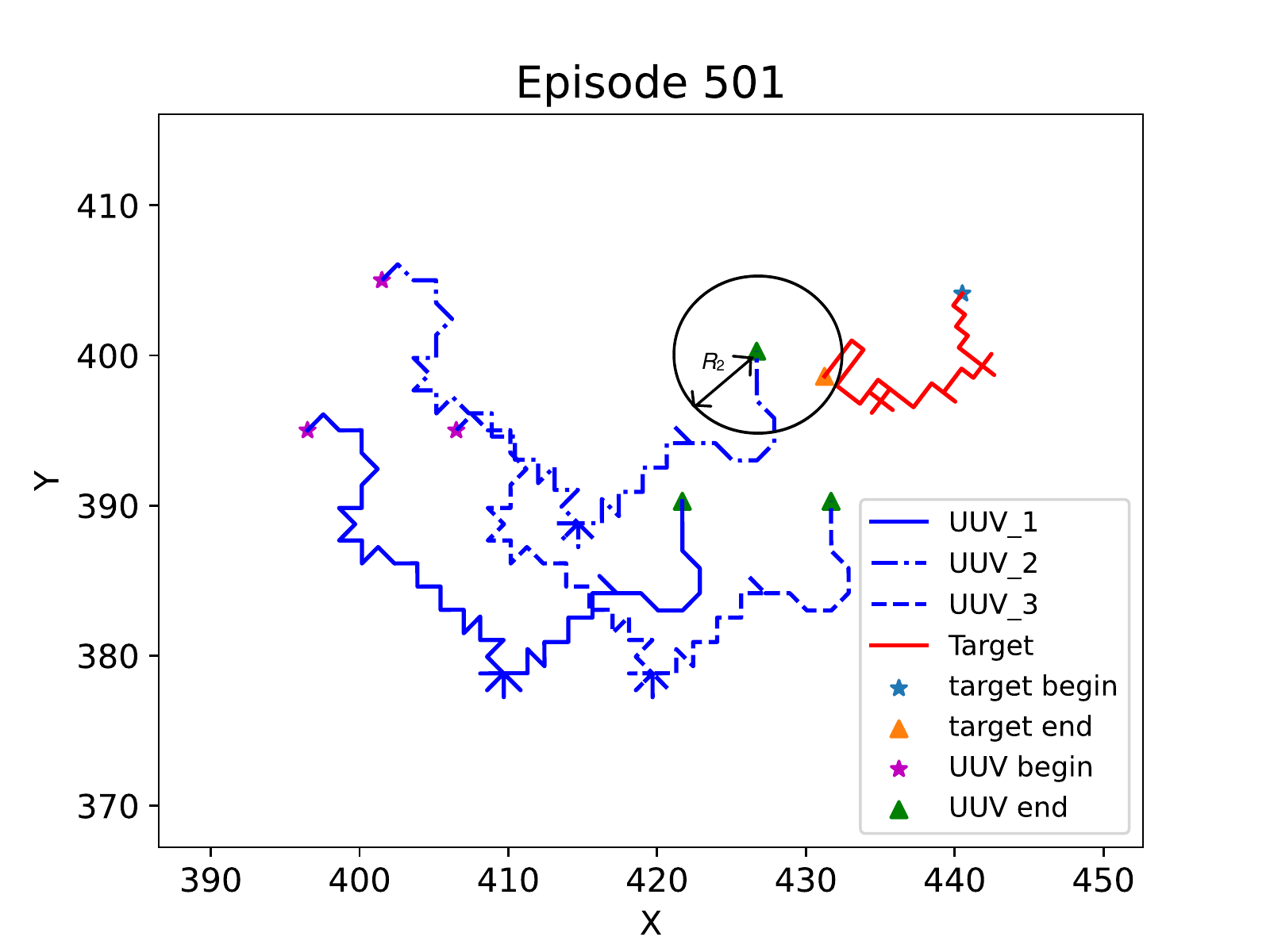}
			\end{minipage}%
		}
		\centering\subfloat[Rewards with various $\bm \tau_d$ when $ \delta \in (0,T_h-t_0)$.] 
		{
			\begin{minipage}[t]{6cm}
				\centering
				\includegraphics[height=4.5cm]{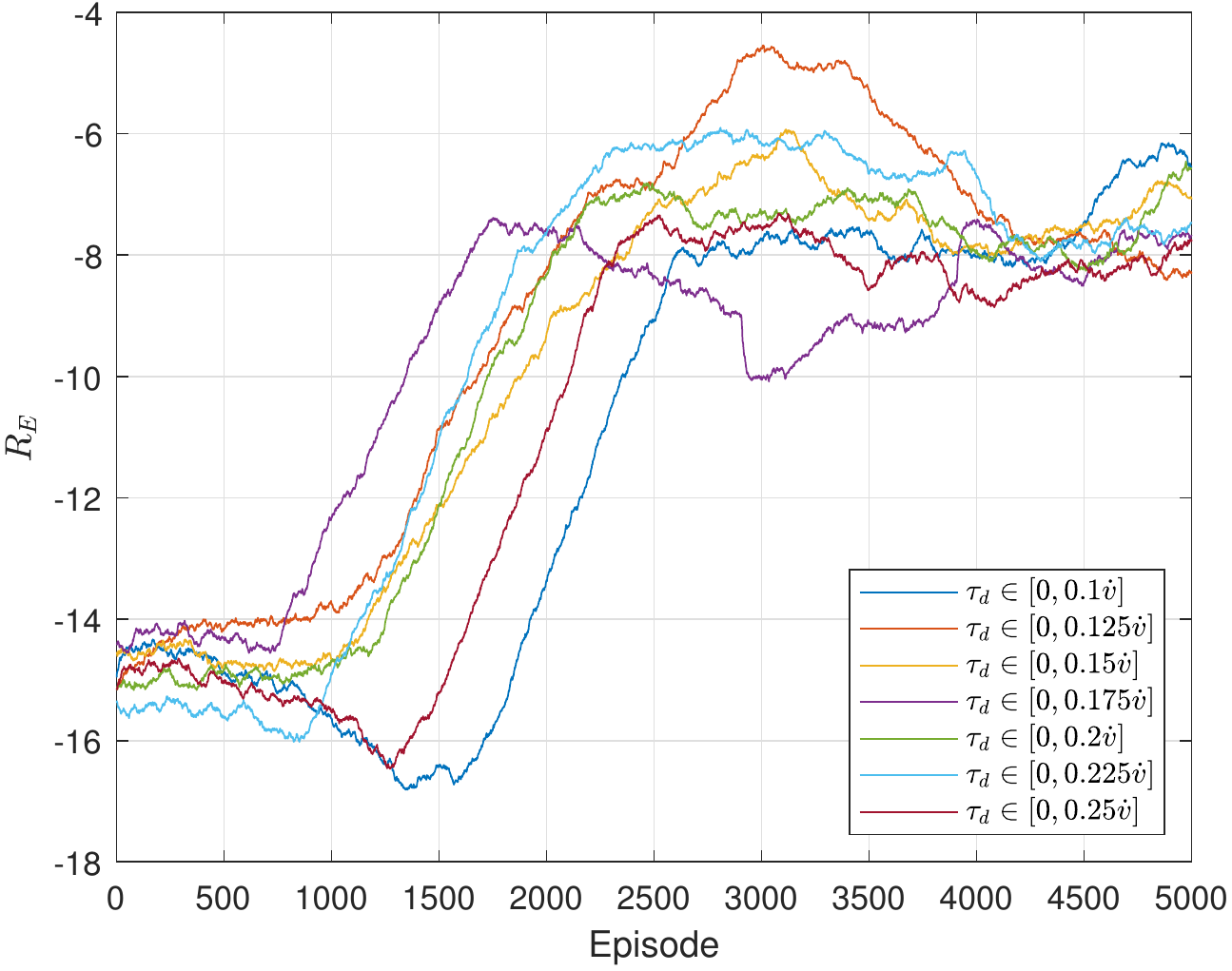}
			\end{minipage}%
		}
		\centering\subfloat[ Consistency test with various $\bm \tau_d$ when $\delta \in (0,T_h-t_0)$.] 
		{
			\begin{minipage}[t]{6cm}
				\centering
				\includegraphics[height=4.5cm]{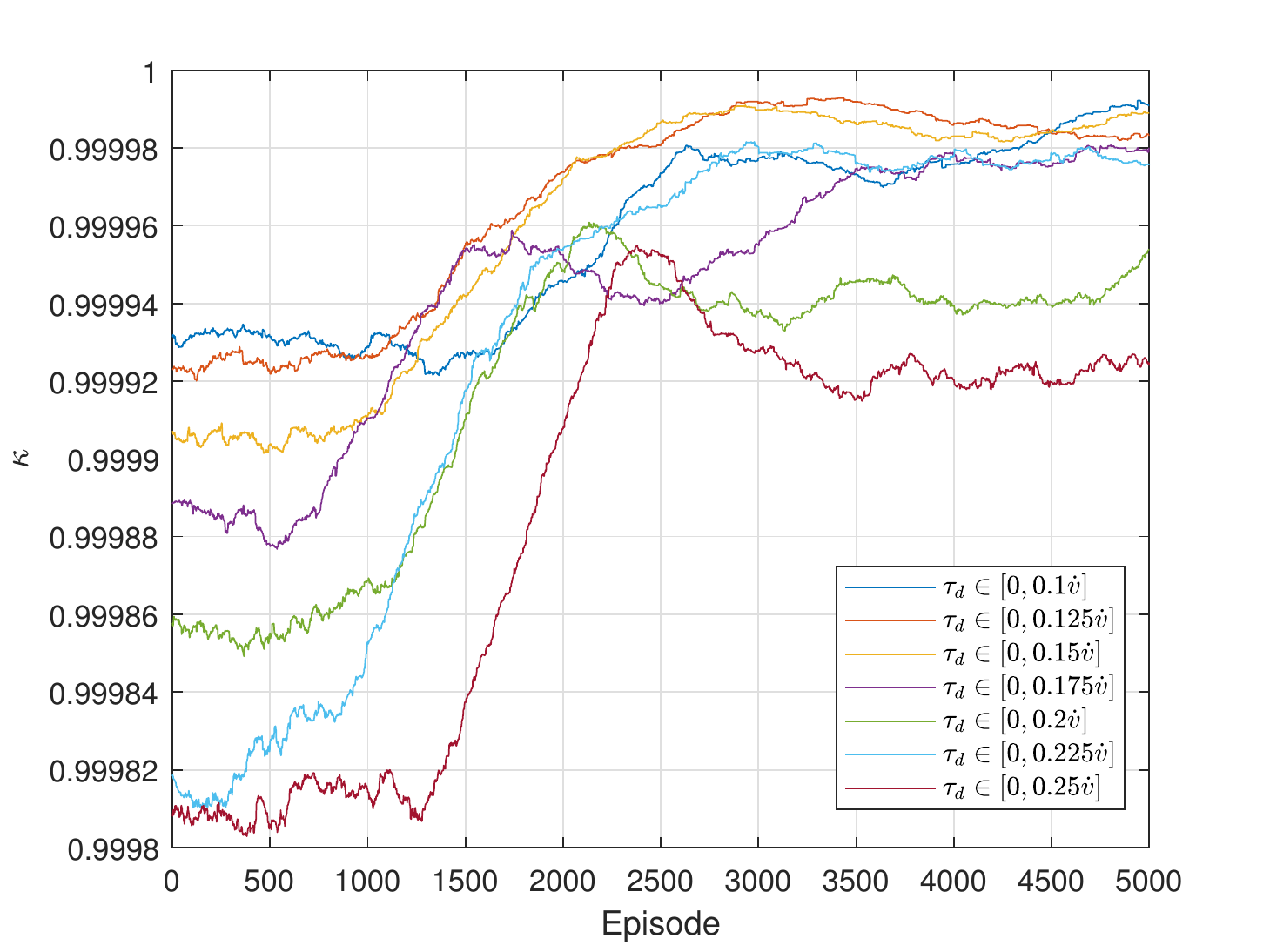}
			\end{minipage}%
		}
		\centering\caption{Underwater differential game between UUVs and the target.} 
	
	\end{figure*}
	
	%

	Here, the DQN method without considering the underwater communication delay is used to compare with Algorithm 1. Moreover, rewards and consistency curves in Fig. 3 are treated by smooth functions. Fig. 3 (a) and Fig. 3 (b) show two underwater target hunting examples when $\delta=0$ and $ \delta \in (0,T_h-t_0)$, respectively. Results validate that  when considering the communication delay, responses of UUVs perform  hysteresis property. By applying the designed target hunting difference game and the modified DQN method in Algorithm 1, we record the total rewards during each episode, as shown in Fig. 3(b) and Fig. 3(e). The conclusion can be drawn that underwater disturbances  have a larger impact on the system when $ \delta \in (0,T_h-t_0)$, while the paths are relatively smoothing when $\delta=0$. Moreover, rewards versus $\tau_d$ converge at different episodes, which reveals that different ranges of disturbances affect convergences and rewards of the system.

	Initially, a single UUV is blind to other UUVs in an unknown environment. However, in order to complete the underwater target hunting task efficiently, UUV needs to adapt the control policy $\bm p_i$ by identifying whether other UUVs have a closer or weaker cooperative relationship.
	Based on the underwater information exchange, there is a simple inference technique to analyze the consistency between UUVs. In each training episode, the pay-off function $P_i(\boldsymbol{p}_i,\boldsymbol{q},\boldsymbol{s}_0)$ can be recorded, while it can reflect the consistency between UUVs on a certain target hunting process \cite{zhang2021coordination}.
	In this work, the Kendall correlation coefficient $\kappa_{i,j}$  is used to obtain the consistency between UUV $i$ and UUV $j$. Thus, the consistency index $\kappa$ of UUVs in $\ell$ episodes can be defined as:
	\begin{equation}
		\begin{aligned}
			\kappa_{i,j}&=\frac{2}{\ell \left( \ell -1 \right)}\sum_{\mathfrak{m}<\mathfrak{n}}{\text{sgn}\left( P_{i\mathfrak{m}}-P_{i\mathfrak{n}} \right) \text{sgn}\left( P_{j\mathfrak{m}}-P_{j\mathfrak{n}} \right)},\\
			\kappa&=\frac{1}{M}\sum\kappa_{i,j},\ i<j, \ \mathfrak{m}\in[0,\ell], \ \mathfrak{n}\in[0,\ell].
		\end{aligned}
	\end{equation}
	
	With the consistency index $\kappa$, Fig. 3 (c) and Fig. 3 (f) show  UUVs perform a good consistency during underwater target hunting tasks, especially when achieving convergence. Furthermore, we can see that consistency curves with $\delta=0$ converge at about 1800 episodes,  while consistency curves with $\delta \in (0,T_h-t_0)$ converge at about 2300. Interestingly, UUVs have a larger $\kappa$ when considering the communication delay. This is mainly because that the hysteresis property leads to a relatively lazy movement, which improves the consistency. Moreover, UUVs suffering from a relatively small range of disturbances, performs better consistency when conducting target hunting task.


	\section{Conclusion}
	In this paper, an underwater target hunting differential game considering communication delay and underwater disturbances has been proposed. In addition, to investigate the optimal controls of UUVs with minimum pay-off function, the Hamiltonian function is used to gain the feedback control policies and modified DQN method further studies the influence of delay and disturbances on target hunting system. Simulation results have revealed the cost and consistency
	of target hunting task versus communication  delay and underwater disturbances. 

	\section*{Acknowledgment}
This work was supported in part by the National Key R$\&$D Program of China (No. 2020YFD0901000),
in part by the National Natural Science Foundation of China (No. 62071268, 61971257, and 62127801),
in part by the Young Elite Scientist Sponsorship Program by CAST (No. 2020QNRC001).
%
\bibliographystyle{IEEEtran}
\bibliography{citationlist}

\end{document}